\input harvmac
\input epsf
\def\Title#1#2{\rightline{#1}\ifx\answ\bigans\nopagenumbers\pageno0\vskip1in
\else\pageno1\vskip.8in\fi \centerline{\titlefont #2}\vskip .5in}

%%%%%%%%%%%%%%%%%%
%
% Figure macros, SBG 5/93
%
\ifx\epsfbox\UnDeFiNeD\message{(NO epsf.tex, FIGURES WILL BE IGNORED)}
\def\figin#1{\vskip2in}% blank space instead
\else\message{(FIGURES WILL BE INCLUDED)}\def\figin#1{#1}%\epsfverbosetrue
\fi
\def\Fig#1{Fig.~\the\figno\xdef#1{Fig.~\the\figno}\global\advance\figno
 by1}
%
%  ifig   usage:
%
%         \ifig\figlabel{caption}{figfile}{vsize}
%
% where vsize is the desired vertical size of the figure in truein
%
\def\ifig#1#2#3#4{
\goodbreak\midinsert
\figin{\centerline{\epsfysize=#4truein\epsfbox{#3}}}
\narrower\narrower\noindent{\footnotefont
{\bf #1:}  #2\par}
\endinsert
}

%
%
%defs
%
\font\ticp=cmcsc10
\def\sq{{\vbox {\hrule height 0.6pt\hbox{\vrule width 0.6pt\hskip 3pt
   \vbox{\vskip 6pt}\hskip 3pt \vrule width 0.6pt}\hrule height 0.6pt}}}
\def\hmn{h_{\mu\nu}}
\def\ajou#1&#2(#3){\ \sl#1\bf#2\rm(19#3)}
\def\jou#1&#2(#3){,\ \sl#1\bf#2\rm(19#3)}

\def\frac#1#2{{#1\over#2}}
\def\Rads{{R}}
\def\cald{{\cal D}}
\def\sqrtg{\sqrt{-g}}
\def\sqrtG{\sqrt{-G}}
\def\calL{{\cal L}}
\def\calG{{\cal G}}
\def\calR{{\cal R}}
\def\dhg{{\hat \Delta}_>}
\def\dhl{{\hat \Delta}_<}
%
%refs
%
\lref\BDL{P.~Binetruy, C.~Deffayet and D.~Langlois, ``Brane cosmological 
evolution in a bulk with cosmological constant,''
hep-th/9905012.}
\lref\RandC{L. Randall, talk at Caltech/USC conference ``String theory at
the millennium,''
http://quark.theory.caltech.edu/people/rahmfeld/Randall/fs1.html.} 
\lref\Kal{N. Kaloper, ``Bent domain walls as brane-worlds,''
hep-th/9905210.}
\lref\Nih{T.~Nihei, ``Inflation in the five-dimensional universe with an 
orbifold extra
dimension,'' hep-ph/9905487.}
\lref\CGKT{C.~Cs\'aki, M.~Graesser, C.~Kolda and J.~Terning, 
``Cosmology of one extra dimension with localized gravity,''
hep-ph/9906513.}
\lref\CGS{J.M.~Cline, C.~Grojean and G.~Servant,
``Cosmological expansion in the presence of extra dimensions,''
hep-ph/9906523.}
\lref\ChFr{D.J.~Chung and K.~Freese,
``Can geodesics in extra dimensions solve the cosmological horizon
problem?''
hep-ph/9906542.}
\lref\CGRT{C.~Cs\'aki, M.~Graesser, L. Randall, and J.~Terning, ``Cosmology
of brane models with radion stabilization,'' hep-ph/9911406.}
\lref\KiKi{H.B.~Kim and H.D.~Kim, ``Inflation and gauge hierarchy in
Randall-Sundrum compactification,'' hep-th/9909053.}
\lref\KKOP{P.~Kanti, I.I.~Kogan, K.A.~Olive and M.~Pospelov, 
``Single brane cosmological solutions with a stable compact extra
dimension,'' 
hep-ph/9909481.}
\lref\BuLu{C.P. Burgess and C.A. Lutken, ``Propagators and effective
 potentials in anti-de Sitter
space,''\ajou Phys. Lett. &153B (85) 137.}
\lref\InOo{T. Inami and H. Ooguri, ``One loop effective potential
 in anti-de Sitter space,''\ajou Prog. Theor. Phys. &73 (85) 1051.}
\lref\DDG{K.R.~Dienes, E.~Dudas and T.~Gherghetta,  
``Invisible axions and large-radius compactifications,''
hep-ph/9908530.}
\lref\DHR{H.~Davoudiasl, J.L.~Hewett and T.G.~Rizzo,  
``Bulk gauge fields in the Randall-Sundrum model,''
hep-ph/9909255.}
\lref\CHR{A. Chamblin, S.W. Hawking, and H.S. Reall, ``Brane-world black
holes,'' hep-th/9909205.}
\lref\RaSui{L. Randall and R. Sundrum, ``A large mass hierarchy from a
small extra dimension,'' hep-ph/9905221\jou Phys. Rev. Lett. &83 (99) 3370.} 
\lref\RaSu{L. Randall and R. Sundrum, ``An alternative to
compactification,'' hep-th/9906064\jou Phys. Rev. Lett. &83 (99) 4690.}
\lref\Viss{M. Visser, ``An exotic class of Kaluza-Klein 
models,''\ajou Phys. Lett. &B159 (85) 22.}
\lref\RuSh{V.A. Rubakov and M.E. Shaposhnikov, 
``Do we live inside a domain wall?,"\ajou Phys. Lett. &125B (83) 136.}
\lref\HoSt{G. Horowitz and A. Strominger, ``Black strings and 
p-branes,''\ajou Nucl. Phys. &B360 (91) 197.}
\lref\Hver{H. Verlinde, ``Holography and compactification,''
hep-th/9906182.}
\lref\MaldUnp{J. Maldacena, unpublished.}
\lref\WittITP{E. Witten, remarks at ITP Santa Barbara Conference ``New
dimensions in field theory and string theory,''
http://www.itp.ucsb.edu/online/susy\_c99/discussion/.}
\lref\GiddITP{S.B. Giddings, talk at ITP Santa Barbara Conference ``New
dimensions in field theory and string theory,''
http://www.itp.ucsb.edu/online/susy\_c99/giddings/.}
\lref\FSS{S.B. Giddings, ``Flat-space scattering and bulk locality in the
AdS/CFT correspondence,'' hep-th/9907129.}
\lref\Mald{J. Maldacena, ``The large-N limit of superconformal field
theories and supergravity,'' hep-th/9711200\jou Adv. Theor. Math. Phys. &2
(98) 231.}
\lref\GKP{S.S. Gubser, I.R. Klebanov, and A.M. Polyakov, ``Gauge theory
correlators from noncritical string theory,'' hep-th/9802109\jou
Phys. Lett. &B248 (98) 105.}
\lref\WittAds{E. Witten, ``Anti-de Sitter space and holography,''
hep-th/9802150\jou Adv. Theor. Math. Phys. &2 (98) 253.}
\lref\Gubs{S.S. Gubser, ``AdS/CFT and gravity,'' hep-th/9912001.}
\lref\EHM{R. Emparan, G.T. Horowitz, and R.C. Myers, ``Exact description of
black holes on branes,'' hep-th/9911043.}
\lref\GaTa{J. Garriga and T. Tanaka, ``Gravity in the brane world,'' 
hep-th/9911055.} 
\lref\BGL{V. Balasubramanian, S.B. Giddings, and A. Lawrence, ``What
do CFTs tell us about anti-de Sitter space-times?'' 
hep-th/9902052, {\sl JHEP} {\bf 9903:001} (1999).}
\lref\Isra{W. Israel, ``Singular hypersurfaces and thin shells in general
relativity,''\ajou Nuovo Cim. &44B (66) 1.}
\lref\MTW{C.W. Misner, K.S. Thorne, and J.A. Wheeler, {\sl Gravitation}
(W.H. Freeman, 1973).}
\lref\LyRa{J. Lykken and L. Randall, ``The shape of gravity,''
hep-th/9908076.}
\lref\HeSk{M. Henningson and K. Skenderis, ``The holographic Weyl
anomaly,'' hep-th/9806087, {\sl JHEP} {\bf 9807:023} (1998).}
\lref\BaKr{V. Balasubramanian and P. Kraus, ``A stress tensor for anti-de
Sitter gravity'' hep-th/9902121\jou  Commun. Math. Phys. &208 (99) 413.}
\lref\BaKrHol{V. Balasubramanian and P. Kraus, ``Space-time and the
holographic renormalization group,'' hep-th/9903190\jou Phys. Rev. Lett.
&83 (99) 3605.}
\lref\GoWi{W.D. Goldberger and M.B. Wise, ``Modulus stabilization with bulk
fields,'' hep-ph/9907447\jou Phys. Rev. Lett. &83 (99) 4922.}
%\lref\CTR{C. Cs\'aki, M. Graesser, L. Randall, and J. Terning, ``Cosmology
%of models with radion stabilization,'' hep-ph/9911406.}
\lref\Newdim{Online proceedings of the ITP Santa Barbara conference 
``New dimensions in field theory and string theory,''
http://online.itp.ucsb.edu/online/susy\_c99/ .}
\lref\DFGK{O. DeWolfe, D.Z. Freedman, S.S. Gubser, A. Karch, ``Modeling 
the fifth-dimension with scalars and gravity,''
hep-th/9909134.}
\Title{\vbox{\baselineskip12pt
%\hbox{Preliminary Draft/Do Not Distribute}
\hbox{hep-th/0002091}\hbox{MIT-CTP-2944}\hbox{PUPT-1913}
}}
{\vbox{\centerline {%Gravitational Dynamics in Brane Reduction Scenarios
Linearized Gravity in Brane Backgrounds}
}}
\centerline{{\ticp Steven B. Giddings},${}^1$\footnote{$^\dagger$}
{Email address:
giddings@physics.ucsb.edu} {\ticp Emanuel Katz},${}^2$\footnote{$^*$}{Email
address: amikatz@mit.edu }
{\ticp and Lisa Randall}${}^{2,3}$\footnote{$^\sharp$}{Email address:
randall@feynman.princeton.edu}  }
\vskip.1in
\centerline{${}^1$ {\sl Department of Physics, University of California}}
%\centerline{\sl University of California}
\centerline{\sl Santa Barbara, CA 93106-9530}
\vskip.1in
\centerline{${}^2$ {\sl Center for Theoretical Physics, MIT}}
%\centerline{\sl Massachusetts Institute of Technology}
\centerline{\sl Cambridge, MA 02139}
\vskip.1in
\centerline{${}^3$ {\sl Joseph Henry Laboratories, Princeton University}}
%\centerline{\sl Princeton University}
\centerline{\sl Princeton, NJ 08543}
\centerline{and}
\centerline{\sl Institute of Theoretical Physics, University of California}
%\centerline{\sl University of California}
\centerline{\sl Santa Barbara, CA 93106}
%\centerline{{\ticp Emanuel Katz, and Lisa Randall}}
%\centerline{\sl MIT}
\bigskip
\centerline{\bf Abstract}

A treatment of linearized gravity is given in the Randall-Sundrum
background.  The graviton propagator is found  in terms of the scalar
propagator, for which an explicit integral expression is provided.
This reduces to the four-dimensional propagator at long distances along the
brane, and provides estimates of subleading corrections.  Asymptotics of the
propagator off the brane yields  exponential falloff of gravitational
fields due to matter on the brane.  This implies that black holes bound to
the brane have a ``pancake''-like shape in the extra dimension, and
indicates validity of a perturbative treatment off the brane.
Some connections with the AdS/CFT correspondence are described.

%\draft
\Date{}

\newsec{Introduction and summary}

It is possible that the observed world is a brane embedded in
a space with more noncompact dimensions.  This proposal was made more
concrete in the scenario advanced in \refs{\RaSui,\RaSu}, where the problem of
recovering four-dimensional gravity was addressed.  (Earlier
work appears in refs.~\refs{\RuSh\Viss-\HoSt}.)   
Further exploration of this scenario has
included investigation of its
cosmology\refs{\BDL\Kal\Nih\CGKT\CGS\ChFr\KiKi\KKOP-\CGRT} and 
phenomenology\refs{\DDG,\DHR}.\foot{For a
recent survey of some of these topics, see also \refs{\Newdim}.}

In order to ``localize" gravity to the brane, ref.~{\RaSu} worked
in an embedding space with a background cosmological constant, with
total action of the form
%\foot{For conventions, see the appendix.}
%
\eqn\action{S=\int d^{5}X \sqrt{-G} ( -\Lambda + M^3 \calR) + \int d^4x
\sqrt{-g} {\cal L}\ .}
$G$ and $X$ are the five-dimensional metric and coordinates, and $g$
and $x$ are the corresponding four-dimensional quantities with $g$ given as
the pullback of the five-dimensional metric to the brane.  $M$ is the
five-dimensional Planck mass, and $\calR$ denotes
the five-dimensional Ricci scalar.  The bulk space
is a piece of anti-de Sitter space, with radius $\Rads =
\sqrt{-12M^3/\Lambda}$, which has metric
\eqn\adsmet{dS^2 = {\Rads^2\over z^2} (dz^2 + dx_4^2)\ .}
The brane can be taken to reside at $z=\Rads$, or in scenarios \refs{\LyRa}
with both
a probe (or ``TeV'') brane and
a Planck brane, this will be the location of the Planck brane.  The horizon
for observers on either brane is at $z=\infty$.

There are a number of outstanding questions with this proposal. 
One very interesting question is what black holes or more general
gravitational fields, {\it e.g.} due to sources on the brane,
look like, both on and  off the brane.
For example, consider a black hole formed from matter on the brane.
{}From the
low-energy perspective of an observer on the brane it should appear like
a more-or-less standard four-dimensional black hole
but one expects a five-dimensional observer to measure a
non-zero transverse thickness.  One can trivially find solutions that a
four-dimensional observer sees as a black hole by replacing $dx^2$ with the
Schwarzschild metric in \adsmet.\foot{Such metrics were independently found
in \refs{\CHR}.}  However, these ``tubular'' 
solutions become singular at the horizon at
$z=\infty$, suggesting that another solution be found.

Another related question concerns the dynamics of gravity.  
It was argued in \RaSu\ that four-dimensional
gravitational dynamics arises from a graviton zero mode bound to the
brane.
Fluctuations in this zero mode correspond to perturbations of the form
\eqn\gravfluct{dS^2 \rightarrow dS^2 + {R^2\over z^2} h_{\mu\nu} dx^\mu
dx^\nu\ .}
where $\hmn$ is a function only of $x$,
\eqn\hzm{\hmn= \hmn(x)\ .}
Computing the lagrangian of such a fluctuation yields 
\eqn\flulag{\calR\sim {z^2} \partial h\partial h\ .}

This and other measures of the curvature 
of the fluctuation generically grow without
bound as $z\rightarrow\infty$.  In particular, if we add a higher power of
the curvature to the action, with small coefficient, as may be induced from
some more fundamental theory of gravity, then generically divergences will
be encountered.  For example,  one easily estimates 
\eqn\rdiv{ \calR_{\mu\nu\lambda\sigma} \calR^{\mu\nu\lambda\sigma} \sim z^4\ }
suggesting that Planck scale effects are important near the horizon.  
This would raise serious questions about the viability of the underlying 
scenario.  These estimates are however incorrect as they neglect the
non-zero modes.  

Yet another question regards corrections to the 4d gravitational effective
theory on the brane.  We'd like to better understand the strength of 
corrections
to Newton's Law and other gravitational formulae; some of the leading
corrections have already been examined via the mode sum\refs{\RaSu,\LyRa}.  
Sufficiently large
corrections could provide experimental tests of or constraints on these
scenarios.

A final point addresses reinterpretation \refs{\Hver,\MaldUnp,\WittITP} of 
these
scenarios within the context of $AdS/CFT$ duality
\refs{\Mald\GKP-\WittAds}.  In this
picture, gravity in the bulk AdS off
the brane can be replaced by ${\cal N}=4$ super-Yang Mills theory
on the brane.  Witten\WittITP\foot{See also \refs{\Gubs}.} has suggested
that gravitational corrections from the bulk 
can be reinterpreted in terms of the loop diagrams
in the SYM theory.

In order to address these questions, 
this paper will give an analysis of linearized gravity in the background of
\RaSu.  We begin in section two with a 
derivation of the propagator for a scalar field
 in the brane background of \RaSu, generalized to $d$+1
dimensions.  
This exhibits much of the
physics with less complication than gravity.  The propagator is the usual
AdS propagator plus a correction term, and can be rewritten, for sources on
the brane, in terms of a zero-mode contribution that produces
$d$-dimensional gravity on the brane plus a correction from the
``Kaluza-Klein'' modes.  For even $d\geq4$ this term produces corrections of
order $(R/r)^{d-2}$ at large distances $r$ from the source.

In section three we perform a linearization of gravity about the $d$+1
dimensional brane background.  For general matter source on the brane, the
brane has non-zero extrinsic curvature, and a consistent linear analysis
requires introduction of coordinates in which the bending of the brane is
manifest, as exhibited in eq.~(3.22).  We outline the derivation of the
graviton propagator, which can be written in terms of the scalar propagator
of section 2.  (For those readers interested primarily in the applications
discussed in the subsequent sections, the results appear in eqs.~(3.23),
(3.24), and (3.26).)  Special simplifying cases include treatment of sources 
restricted
to the Planck brane, or living on a probe brane in the bulk.

In section four we discuss the asymptotics and physics of the resulting
propagator.  Linearized gravity on the Planck brane corresponds to 
$d$-dimensional
linearized gravity, plus correction terms from Kaluza-Klein modes.  As in
the scalar case, these yield large-$r$ corrections suppressed by
$(R/r)^{d-2}$ 
for even $d\geq4$ and by $R/r$ for $d=3$, in agreement with \EHM.  
We also discuss corrections in the
probe-brane scenario of \LyRa.  
We then find the falloff in the
gravitational
potential off the brane and thus deduce the shape in the extra dimensions 
of black holes bound to
the brane or of more
general gravitational fields.
In particular, we
find that black holes have a transverse size that grows with mass like
$\log m$, compared to the usual result $m^{1/d-3}$ along the brane.  Thus
black holes have a pancake-like shape.  We also check consistency of the
linearized approximation, and check that higher-order curvature terms in
the action in fact 
do not lead to large corrections, as the na\"\i ve analysis of the
zero mode would indicate.  

Finally, section five contains some comments on the connection with the
AdS/CFT correspondence.  An extension of the Maldacena
conjecture{\refs\Mald} may enable one to replace the five-dimensional
physics off the brane by a suitably regulated large-$N$ gauge
theory\refs{\Hver,\MaldUnp,\WittITP}.  In particular, we discuss how this
picture produces four-dimensional gravity plus  correction terms 
like those mentioned above.

During the period during which this work has been completed, several related
works have appeared.  Ref.~\refs{\EHM} has found exact solutions for black 
holes
bound to a brane in a 2+1 dimensional version of this scenario, and in that
case independently discovered the shape of black holes (which
in fact easily follows from the earlier estimates of \refs{\CHR}).
Ref.~\refs{\GaTa} has also outlined aspects of 
a linear analysis of gravity, and in
particular emphasized the importance of the bending of the brane.  Recent
comments on the relation with the AdS/CFT correspondence were made by
Witten \refs{\WittITP}, with further elaboration in \refs{\Gubs}.
Preliminary presentations of some of our results can be found in
\refs{\GiddITP,\RandC}.  

\newsec{The massless scalar propagator}

Much of the physics of linearized gravity in the scenario of \RaSu\ is
actually found in the simpler case of a minimally coupled scalar field.  
Because
of this, and because the scalar propagator is needed in order to compute
the graviton propagator, this section will focus on computing the scalar
Green function.

In much of this paper we will work with the generalization to a
$d$+1 dimensional theory with a brane of codimension one.  The scalar
action, with source terms, takes the form
\eqn\scalact{S= \int d^{d+1} X \sqrtG \left[ -{1\over 2}(\nabla\phi)^2 +
J(X)\phi(X)\right]\ .}
Here we work in the brane background of \RaSu; for $z>R$ the metric is the
$d+1$-dimensional AdS metric, 
\eqn\Adsmet{dS^2 = {{{\Rads}^2\over{z^2}}({dz^2} + {dx_d^2})}  .}
Without loss of generality the brane can be located at $z=R$.  
This problem serves as a toy-model for gravity; for
a given source $J(X)$, the resulting field $\phi(X)$ is analogous to the
gravitational field of a fixed matter source.  The field is given in
terms of the scalar Green function, obeying
\eqn\branegf{
\sq \Delta_{d+1} (X,X^{\prime}) = {\delta^{d+1}(X - X^{\prime})\over {\sqrt {-
G}}}\ .
}
Analogously to the boundary conditions that we will find on the
gravitational field, the scalar boundary conditions are taken to be Neumann,
\eqn\branebc{
\partial_z \Delta_{d+1} (x,x^{\prime})_{\vert_{z=R}} = 0 \ ;
}
these can be interpreted as resulting from the orbifold boundary conditions
at the brane, or alternately as due to the energy density on the wall.
The scalar field has a bound zero mode $\phi=\phi(x)$ analogous to that of
gravity.  

In order to solve \branegf, 
we first reduce the problem to solving an ordinary differential equation
via a Fourier transform in the $d$ dimensions along the wall,
\eqn\fourier{\Delta_{d+1}(x,z;x',z') = \int {d^d p \over 
(2\pi)^d}e^{ip(x-x')} \Delta_p(z,z').}
The Fourier component must then satisfy the equation,
\eqn\plapl{{z^2 \over R^2}({\partial_z}^2 - {d-1 \over z}\partial_z
-p^2)\Delta_p(z,z') = \left({z \over R}\right)^{d+1}
\delta(z-z')\ .}
Making the definitions $\Delta_p = ({zz' \over R^2})^{d\over 2}
{\hat \Delta}_p$
and 
\eqn\Qdef{q^2=-p^2\ ,}
this becomes
\eqn\hatlapl{\left(z^2\partial_z^2 + z\partial_z +q^2z^2 - {d^2\over 4}\right) 
\hat{\Delta}_p(z,z')=Rz\delta(z-
z').}
For $z \neq z'$, the equation admits as its two independent solutions the 
Bessel
functions
$J_{d \over 2}(qz)$ and $Y_{d \over 2}({q}z)$. 
Hence, the solution for $z<z'$ and for $z>z'$ must be linear
combinations $\dhl(z,z')$, $\dhg(z,z')$ 
of these functions.  Eq.~\hatlapl\ then implies matching conditions at 
$z=z'$:
\eqn\matching{\eqalign{
\dhl|_{z=z'} = &  \dhg|_{z=z'} \cr
\partial_z (\dhg-\dhl)|_{z=z'} =& {R\over z'}\ .}}
%
%The propagator will be symmetric under interchange of $z$ and $z'$.

We begin with the Green function for $z<z'$.
%
%\eqn\Gansatz{\dhl =\left(A_<(z')J_{d \over 2}(qz) +
%B_<(z')Y_{d \over 2}(qz)\right)\ .}
%
%where we have let $q^2 = -p^2$.
The boundary condition \branebc\ translates to
%
%\eqn\boundhat{\partial_z \hat{G}_{p,z<z'}|_{z=R} = -{d-1 \over 2R} 
%\hat{G}_{p,z<z'}|_{z=R}.}
%
\eqn\boundhat{\partial_z\left[z^{d/2} \dhl\right]|_{{z=R}}= 0\ .}
This has solution
%
%\eqn\bcond{A_<J_{{d\over 2}-1}(qR) +
%B_<Y_{{d\over 2}-1}(qR) = 0.}
%
\eqn\reparam{\eqalign{
\dhl = &
A(z')\left[Y_{{d \over 2}-1}(qR)J_{d \over 2}(qz)-J_{{d \over 
2}-1}(qR)Y_{d \over 2}(qz)\right]
\cr
 =&
iA(z')\left[J_{{d\over 2}-1}(qR)H_{d \over 2}^{(1)}(qz)-H_{{d 
\over 2}-1}^{(1)}(qR)J_{d \over 2}(qz)\right],}}
where $H^{(1)} = J+iY$ is the first Hankel function.

Next, consider the region $z>z'$.  The boundary conditions at the horizon
$z=\infty$ are analogous to the Hartle-Hawking boundary conditions and are 
inferred by demanding that positive frequency waves be ingoing there,
implying\refs{\BGL}
%The presence of a horizon at
%$z=\infty$, requires that for $t'>t$ there be no propagation from
%$z=\infty$ to any $z'$.  Hence, we must choose the linear combination,
%which at infinity, corresponds to propagation into the horizon only 
%Thus,
%
\eqn\Gansatztwo{ \dhg = B(z')H_{d \over 2}^{(1)}(qz)\ .}
%
%where $H_{d \over 2}^{(1)}(qz)$ is the first Hankel function.
%Letting 
%
%\eqn\reparam{\eqalign{
%\dhl = &
%A(z')\left[Y_{{d \over 2}-1}(qR)J_{d \over 2}(qz)-J_{{d \over 
%2}-1}(qR)Y_{d \over 2}(qz)\right]
%\cr
% =&
%iA(z')\left[J_{{d\over 2}-1}(qR)H_{d \over 2}^{(1)}(qz)-H_{{d 
%\over 2}-1}^{(1)}(qR)J_{d \over 2}(qz)\right],}}
%

The matching conditions \matching\ between the regions then become
\eqn\matchcond{\eqalign{
& iA(z')\left[J_{{d\over 2}-1}(qR)H_{d \over 2}^{(1)}(qz') - H_{{d
\over 2}-1}^{(1)}(qR)J_{d \over 2}(qz')\right]
= B(z') H_{d \over 2}^{(1)}(qz'), \cr
& B(z') H_{d \over 2}^{(1)\prime}(qz')
-
iA(z') \left[J_{{d \over 2}-1}(qR)H_{d \over 2}^{(1)\prime}(qz')
-H_{{d \over 2}-1}^{(1)}(qR)J_{d \over 2}'(qz')\right]
= {R \over {qz'}}\ .}}
The solution to these gives
\eqn\ghat{
\hat{\Delta}_p = i
{\pi \over 2}R\left[J_{{d \over 2}-1}(qR)H_{d \over
2}^{(1)}(qz_<)-H_{{d\over 2}-1}^{(1)}(qR)J_{d \over 2}(qz_<)\right]{H_{d
\over 2}^{(1)}(qz_>) \over H_{{d \over 2}-1}^{(1)}(qR)},}
where $z_>$ ($z_<$) denotes the greater (lesser) of $z$ and $z'$.
This leads to
the final expression for the scalar propagator: 
\eqn\finalG{\eqalign{
\Delta_{d+1}(x,z;x',z')& ={i\pi\over 2R^{d-1}}(zz')^{d \over 2} \int  {d^d p 
\over
(2\pi)^d}e^{ip(x-x')} \cr
& \times \left[{J_{{d \over 2}-1}(qR) \over H_{{d \over 2}-1}^{(1)}(qR)}
H_{d \over 2}^{(1)}(qz) H_{d \over 2}^{(1)}(qz') 
-J_{d \over 2}(qz_<)H_{d \over 2}^{(1)}(qz_>)\right]\ .}}
We note that the second term is nothing but the ordinary massless scalar
propagator in $AdS_{d+1}$. 

A case that will be of particular interest in subsequent sections is that
where one of the arguments of $\Delta_{d+1}$ is on the Planck brane, at
$z=R$.  In this case, the propagator is easily shown to reduce to
\eqn\onbrane{ \Delta_{d+1}(x,z;x',R) = \left({z\over R}\right)^{d/2} \int
{d^dp\over  (2\pi)^d} e^{ip(x-x')} {1\over q} {H_{d\over 2}^{(1)}(qz) \over 
H_{{d\over 2}-1}^{(1)}(qR)}\ .} 
For both points at $z=R$, a Bessel recursion relation gives a more
suggestive result:
\eqn\braneGF{  \Delta_{d+1}(x,R;x',R) = \int
{d^dp \over (2\pi)^d} e^{ip(x-x')} \left[ {d-2\over q^2 R}  - {1\over q}  
{H_{{d\over
2}-2}^{(1)}(qR) \over 
H_{{d\over 2}-1}^{(1)}(qR)}\right] \ .}
This can clearly be separated into  the standard $d$-dimensional scalar
propagator $\Delta_d$, with
\eqn\Gddef{
\partial_{\mu}\partial^{\mu}\Delta_{d}(x,x')=\delta^{d}(x-x')\  ,
}
which is produced by the zero-mode, 
plus a piece due to exchange of Kaluza-Klein states:
\eqn\propsep{  \Delta_{d+1}(x,R;x',R) = {d-2\over R} \Delta_d(x,x') +
\Delta_{KK}(x,x')\ .}
Here
\eqn\kkcontrib{ \Delta_{KK}(x,x') = -\int
{d^dp \over  (2\pi)^d} e^{ip(x-x')}{1\over q}  {H_{{d\over
2}-2}^{(1)}(qR) \over 
H_{{d\over 2}-1}^{(1)}(qR)}\ .}
Note that for $d=3$, this gives the very simple result
\eqn\kkthree{\Delta_{KK} = i\int {d^3p \over (2\pi)^3 } {e^{ip(x-x')}\over q} =
{1\over 2\pi^2 |x-x'|^2}  \ .}
(One must however be careful in treating the gravitational field 
as a perturbation in
this case; recall that in $d=3$ the potential is logarithmic, which 
is in a sense not a
small perturbation on a Minkowski background.)

The effective action for exchange of $\phi$ fields between two sources as
in \scalact\ is given by the usual quadratic expression involving this
propagator.  While we'll postpone general discussion of the 
asymptotics of these expressions for large or small $x$ or $z$ until we
discuss the graviton propagator, it is worth noting that at large
distances on the brane, $|x-x'|\gg R$, the zero mode piece dominates and we
reproduce the standard effective action for $d$-dimensional scalar
exchange, plus subleading corrections from the Kaluza-Klein part.

\newsec{Linearized gravity}
\subsec{General matter source}

We next turn to a treatment of linearized gravity in the context of brane
reduction.  We again work with a $d+1$-dimensional theory,
with action
\eqn\dact{
S = \int{d^{d+1}}X\sqrt{-G}({M^{d-1}}\calR-\Lambda+{\cal L}_{\rm matter}) - 
\int{d^{d}{x}}
\sqrt{-g}\tau
\ .}
Here ${\cal L}_{\rm matter}$ may include both matter in the bulk and
restricted to the brane.  We have explicitly separated out the brane tension
$\tau$ from the matter lagrangian.  Away from the Planck brane, the vacuum
solution is $d+1$-dimensional $AdS$ space, with metric \Adsmet.
%
%\eqn\Adsmet{ds^2 = {{{\Rads}^2\over{z^2}}({ds^2} + {dx^2})}  .}
%
The $AdS$ radius is determined by solving Einstein's equations.  Denote the
Einstein tensor by $\calG_{IJ}$; these then take the form
\eqn\Eineq{
{{\cal G}_{IJ}}={1\over{2M^{d-1}}} \left[{T_{IJ}}-{\Lambda}{G_{IJ}}-
{\tau}{P_{IJ}}{{\delta({X^{d+1}}-{{X^{d+1}}(x)})}}\right]}
where ${P}_{IJ}$ is the projection operator parallel to the brane, 
given in terms of unit normal ${n}^I$ as 
\eqn\Proj
{P_{IJ} = G_{IJ}-{n_I}{n_J}\ ,}
and ${X^{d+1}}(x)$ gives the position of the 
brane in terms of its intrinsic coordinates $x^\mu$.
Off the brane, \Eineq\ gives
\eqn\adsrad{
{\Rads} = \sqrt{-d(d-1)M^{d-1}\over{\Lambda}}.}
As in \refs{\RaSu}, the brane tension is fine-tuned to give a 
Poincare-invariant solution with symmetric (orbifold) boundary conditions 
about the brane; this condition is
\eqn\tens{
\tau={4(d-1){M^{d-1}}\over{\Rads}}.}
The location of the brane is arbitrary; we take it to be $z={\Rads}$.

The rest of this subsection will focus on deriving the linearized
gravitational field due to an arbitrary source; the
results are presented
in eqs.~(3.23), (3.24), and (3.26) for readers not wishing to follow the
details of the derivation.  As we'll see, maintaining the linearized 
approximation
requires choosing a gauge in which the brane is bent, with displacement
given in eq.~(3.22).

It is often easier to work with the coordinate $y$, defined by
\eqn\ydef{
z=\Rads{e^{y/\Rads}}  ,}
in which the $AdS$ metric takes the form
\eqn\yAdS{
{ds^2}={dy^2}+{e^{{-2|y|}/\Rads}}{\eta_{\mu\nu}}{dx^{\mu}}{dx^{\nu}}   ;}
the brane is at $y=0$, and we have written the solution in a form valid for
all $y$.

It is convenient to describe fluctuations about \yAdS\ in Riemann normal
(or hypersurface orthogonal) coordinates, which can be locally defined for
an arbitrary spacetime metric which then takes the form
\eqn\GNorm{
{ds^2}={dy^2}+{g_{\mu\nu}}(x,y){dx^\mu}{dx^\nu}   .}
The coordinate $y$ picks out a preferred family of hypersurfaces, 
$y=const$.  Such coordinates are not unique; the choice of a base
hypersurface on which they are constructed is arbitrary.
This base hypersurface may be taken to be the brane, but 
later another choice will be convenient.

\ifig{\Fig\One}{A generic deformation of the base surface leads to a
redefinition of gaussian normal coordinates.}{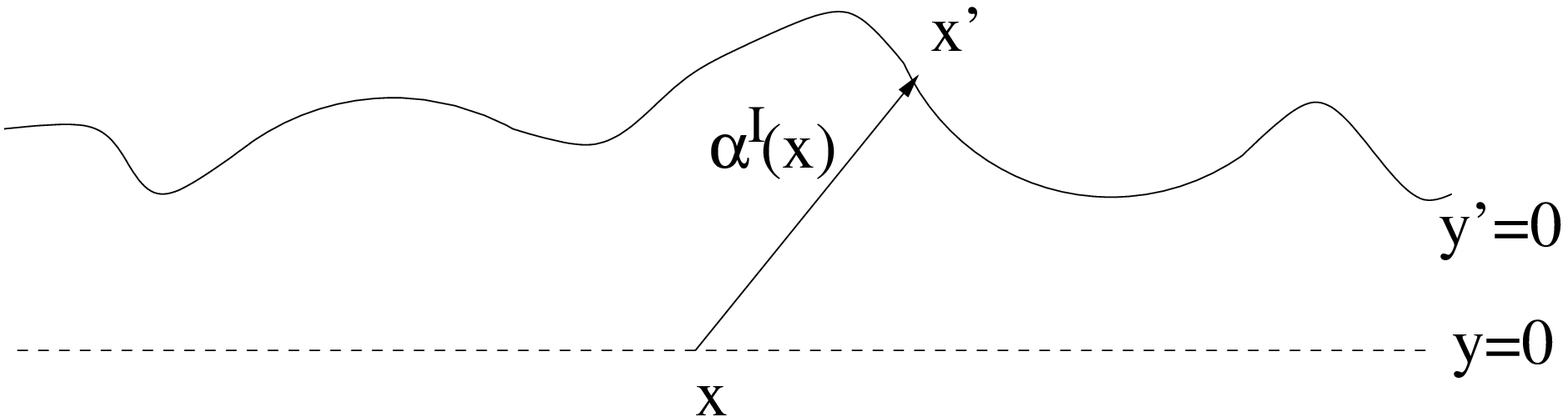}{1.1}

In the case where the coordinates are based on the brane, 
small fluctuations in the metric can be represented as
\eqn\metfluct{
{ds^2}={dy^2}+{e^{-2|y|/\Rads}}[{\eta_{\mu\nu}}+{h_{\mu\nu}(x,y)]
dx^{\mu}dx^{\nu}}
\ .}
Parameterize  a deformation of the coordinates corresponding to changing the 
base hypersurface (see {fig.~1}) by
\eqn\hypdef{
y'=y-{\alpha^y}(x,y)\ ;\ 
{x^\mu}'=  {x^\mu}'(x,y) = {x^\mu}-{\alpha^\mu}(x,y)\ }
and consider a small deformation in the sense that $\alpha^y$ is small.
Working at $y>0$, the condition that the metric takes Gaussian normal form
\GNorm\ in the new coordinates is\foot{Here we have suppressed a subtlety
that
arises in trying to
write a similar expression valid on both sides of the brane due to
the discontinuity in the change of coordinates \hypdef\ across the brane}
\eqn\Gcond{\eqalign{
2\partial_{y'} \alpha^y +g_{\mu\nu} {\partial x^\mu\over  \partial y'} 
{\partial x^\nu\over  \partial y'} &=0 \cr   
\partial_{\mu'} \alpha^y + g_{\mu\nu} {\partial x^\mu\over  \partial y'} 
{\partial x^\nu\over  \partial x^{\mu'}} &=0\ .
}}
In the background metric \yAdS, at $y\roughly< R$ this has general solution  
in terms of arbitrary small functions of $x$:
\eqn\coordsol{
\eqalign {
\alpha^y &= \alpha^y(x)\cr
\alpha^{\mu}(x,y)
&= - {R\over 2} \, e^{2y/\Rads}\partial^{\mu}\alpha^y(x^{\nu})
+ \beta^{\mu}(x)\, .\cr}
}
The corresponding small gauge transformation in the fluctuation $h_{\mu\nu}$ is
\eqn\gaugeXm{
h_{\mu \nu} \rightarrow h^{\prime}_{\mu \nu} = h_{\mu \nu} +
\partial_{\mu}\alpha_{\nu} + \partial_{\nu}\alpha_{\mu} - {2\alpha^y\over
R}\, \eta_{\mu \nu}\, .
}

It is straightforward to expand Einstein's equations \Eineq\ in the fluctuation
$h_{\mu\nu}$; at linear order the result is
\eqn\yy{
2{\cal G}_y^{(1)y} = - ^{(d)}\calR - {(d-1)\over R} \, \partial_yh\epsilon(y) =
{T^y_y\over M^{d-1}}
}
\eqn\muy{
2{\cal G}_\mu^{(1)y} = \partial_y\partial^{\nu} (h_{\mu \nu} - h\eta_{\mu 
\nu}) =
{T^y_{\mu}\over M^{d-1}}
}
\eqn\munu{{\cal G}_\nu^{(1)\mu}=
{}^{(d)}{\cal G}^{\mu}_{\nu} + {d\epsilon(y)\over 2R}\, \partial_y
(h^{\mu}_{\nu} - \delta^{\mu}_{\nu}h) - {1\over 2}\, \partial^2_y
(h^{\mu}_{\nu} - \delta^{\mu}_{\nu}h) = { T^{\mu}_{\nu} {e^{2\vert
y\vert/\Rads}\over 
2M^{d-1} }}}
where ${\cal G}^{(1)}$ denotes the linear part of the Einstein tensor.
In these and subsequent equations, indices on ``small" quantities
$h_{\mu\nu}$, $T_{\mu\nu}$, etc. are raised and lowered with
$\eta_{\mu\nu}$.  ${}^{(d)}\calR$ and ${}^{(d)}{\cal G}^{\mu}_{\nu}$ represent
the curvatures of the induced $d$-dimensional metric of \GNorm\
(which include the conformal factor and will
subsequently be expanded in $h$), and we have used the definitions
\eqn\Edef{\epsilon(y)=\cases{1 &if $y > 0$;\cr
                   -1 &if $y< 0$ \cr}}
and $h=h^{\mu}_\mu$.

Boundary conditions on $h_{\mu\nu}$ at the brane are readily deduced by
integrating the equation \munu\ from just below to just above the brane --
resulting in the Israel matching conditions \refs{\Isra} -- and
enforcing symmetry under $y\to-y$.  If the energy momentum tensor includes
a contribution from matter on the brane,
\eqn\braneT{
T_{\mu\nu}^{\rm brane} = S_{\mu\nu}(x)\delta(y)\, , \hskip.2in T_{yy}^{\rm
brane} = T_{y\mu}^{\rm brane} = 0}
then we find 
\eqn\bcond{
\partial_y(h_{\mu\nu} - \eta_{\mu\nu}h )_{\vert y=0} = - {S_{\mu\nu}(x)\over
2M^{d-1} }\, .}

The first step in solving Einstein's equations \yy-\munu\ is to eliminate
${{}^{(d)}{\calR}}$ between the $(yy)$ and $(\mu\mu)$ equations, resulting in 
an
equation for $h$ alone, (working on the $y>0$ side
of the brane)
\eqn\dhdy{
\partial_y\left( e^{-2y/\Rads}\partial_y h\right) = 
{1\over (d-1)M^{d-1}}  \biggl[ T_{\mu}^{\mu} - (d-2)e^{-2 y/R}\, T^y_y\biggr] 
\ .
}
Conservation of $T$ allows this to be rewritten
\eqn\hsoln{
\partial_y\left[ e^{-2y/\Rads}\left(\partial_y h+ {R\over (d-1)M^{d-1}}
T_{yy}
\right)\right] = - {R\over (d-1)M^{d-1}} \partial^{\mu} T_{\mu
y}\ .}
This can be integrated with initial condition supplied by the trace of
\bcond. 
There is however an apparent problem if 
$S^{\mu}_{\mu}\neq 0$    or      $\partial^{\mu}T_{\mu y}\neq 0$:
the resulting $h$ grows 
exponentially, leading to failure of the linear approximation.

\ifig{\Fig\Two}{In the presence of matter on the brane, the brane is bent
with respect to a coordinate system that is ``straight'' with respect to the 
horizon.}{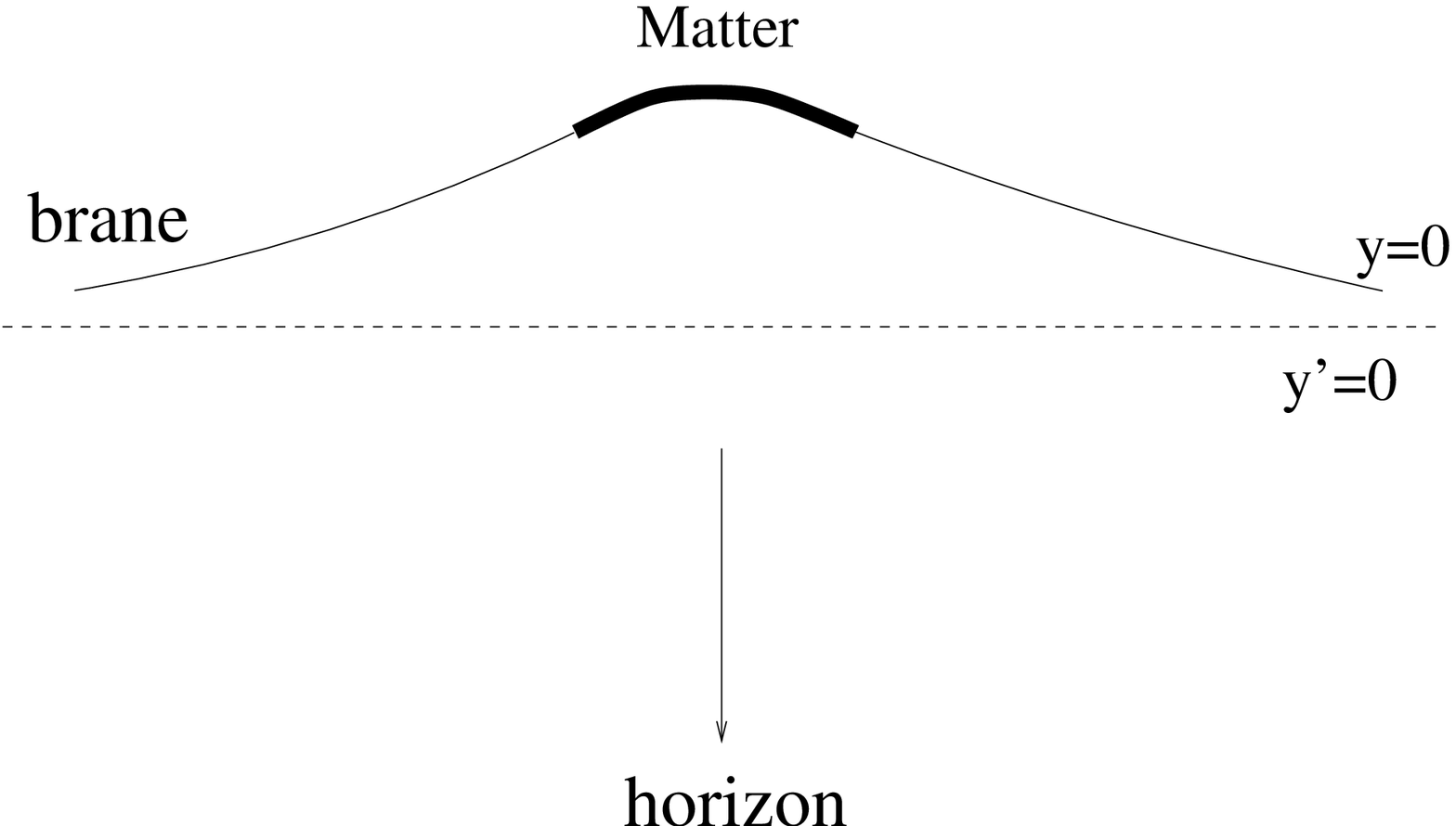}{2.0}

Fortunately this is a gauge artifact, resulting from basing
coordinates on the brane.  Indeed, non-vanishing
${{S_{\mu\nu}}(x)}$
produces extra extrinsic curvature on the brane; to avoid pathological
growth in perturbations one should choose coordinates that are
straight with respect to the horizon.  
In this coordinate system, the
brane appears bent, as was pointed out in \refs{\GaTa}. 
(See fig.~2).  
For simplicity consider the case where all matter is
localized to $y<y_m$, for some $y_m$.  First, the exponential growth in $h$
due to
the initial condition \bcond\ can be eliminated by a coordinate
transformation of the form \coordsol.  We may then integrate up in $y$
until we encounter another source for this growth due to $T_{\mu y}$ on the
RHS of \hsoln, and kill that by again performing a small deformation of the
Gaussian-normal slicing.  We can proceed iteratively at increasing $y$ in
this fashion, with net result that the exponentially growing
part of $h$ can be eliminated by a general slice deformation satisfying
\eqn\bendf{
\partial_{\mu}\partial^{\mu}\alpha^y(x) = {1\over 2(d-1)M^{d-1}} \, \biggl[
{S^{\mu}_{\mu}(x)\over 2}+RT_{yy}(0) - R \int^{y_m}_0 dy \partial^{\mu} T_{\mu
y}\biggr] \ ;
}
in this equation and the remainder of the section, we work in the region on
the $y>0$ side of the brane.
In particular, consider the case ${\partial^\mu}{T_{\mu y}}=T_{yy}(0)=0$; the
solution to \bendf\ then explicitly gives the bending of the brane due to
massive matter on the brane.

In order to solve Einstein's equations we'll therefore work 
with the metric fluctuation $h'_{\mu\nu}$ in this gauge, 
which for small coordinate transformations is
given by \coordsol, \gaugeXm, and \bendf\
(the spatial piece ${\beta^\mu}(x)$ is still arbitrary).
Eq.~\hsoln\ has first integral
\eqn\hpsol{
\partial_y h^{\prime} = - {\Rads\over (d-1)M^{d-1}} \biggl[ T_{yy}(y) -
e^{2y/\Rads} \, \int^{y_m}_y\, dy \partial^{\mu} T_{\mu y}\biggr]\,  
}
and can be solved by quadrature, up to the boundary conditions at $y=0$.
Eq.~\muy\ is then
\eqn\muysol{
\partial_y\partial^{\nu} h^{\prime}_{\mu\nu} = \partial_y\partial_{\mu}
h^{\prime} + {T_{\mu}^y\over M^{d-1}}
}
and can be integrated, again given the boundary conditions, 
to give the longitudinal piece of $h'_{\mu\nu}$.  
Finally, linearizing ${}^{(d)}\calG^{\mu}_{\nu}$ in \munu\ and defining
\eqn\bardel{
{\bar{h}_{\mu\nu}} = h_{\mu\nu} - {1\over 2} \eta_{\mu\nu} h
}
gives
\eqn\munueq{
   \eqalign {
\sq{ \bar{h'}_{\mu\nu} }
&= e^{2y/\Rads}
(-\eta_{\mu\nu}\partial^{\lambda}\partial^{\sigma}{\bar{h}'_{\lambda\sigma}} +
\partial^{\lambda}\partial_{\mu}{\bar{h}^{\prime}}_{\nu\lambda}
+\partial^{\lambda}\partial_{\nu}{\bar{h}^{\prime}}_{\mu\lambda} )\cr
&+ { \eta_{\mu\nu}\over 2 }\, e^{yd/R} \, \partial_y
(e^{-yd/R}\partial_y{h'})\cr
&- { e^{2y/R}\over M^{d-1}} \, T_{\mu\nu}\, .\cr}
}
In this expression, $\sq$ is the {\it scalar} anti-de Sitter
laplacian.  All quantities on the right-hand side of \munueq\ are
known, so ${h'_{\mu\nu}}$ is determined in terms of the scalar Green
function for the brane background, found in the preceding section. 
The metric deformation itself 
is given by trace-reversing,
\eqn\Trrev{
h_{\mu\nu} = {\bar{h}_{\mu\nu}} - {{\bar{h}\over d-2}} \, 
\eta_{\mu\nu}\, .
}

Note, however, that \munueq\ also suffers potential difficulty
from exponentially growing sources.  By \hpsol\ we see that for a
bounded matter distribution, the trouble only lies in the terms
involving $\partial^{\nu}{\bar{h}'}_{\mu\nu}$.  Note also that outside
matter ${\partial_y}\partial^{\nu}{ \bar{h}'_{\mu\nu} }=0$, from \muysol\
and \hpsol.  Therefore, the remaining gauge invariance
$\beta^{\mu}(x)$ in \coordsol\ can be used to set these contributions to
zero just outside the matter distribution,
\eqn\xgauge{\partial^{\nu}\bar{h}'_{\mu\nu}|_{y=y_m}=0\ ,}
and the same holds for all $y>y_m$, eliminating the difficulty. Eq.~\munueq\ 
can then be solved for $h'_{\mu\nu}$ using the scalar Green
function $\Delta_{d+1}(X,X')$, given in eq.~\finalG.  

This will give an
explicit (but somewhat complicated) formula for the gravitational Green
function, defined in general by
\eqn\gravg{ h_{IJ}(X) = {1\over M^{d-1}} \int d^{d+1} X \sqrtG
\Delta_{IJ}^{KL}(X,X') T_{KL}(X')\ ,}
and which can be read off in this gauge from \hpsol, \muysol, and \munueq.
In order to better understand these results, the following two subsections
will treat two special cases.

\subsec{Matter source on the brane}

Consider the case where the only energy-momentum is on the brane.  
In terms of the flat-space Green function $\Delta_d$, 
%
%\eqn\Gddef{
%\partial_{\mu}\partial^{\mu}\Delta_{d}(x,x')=\delta^{d}(x-x')  ,
%}
%
\bendf\ determines a brane-bending function of the form
\eqn\bbend{
\alpha^y(x) = {1\over 4(d-1)M^{d-1}} \int\, d^{d}x' \,
\Delta_d(x,x')\, S^{\mu}_{\mu} (x')\, .
}
Eq.~\hpsol\ and \muysol\ then imply
\eqn\heq{
\partial_yh'=0=\partial_y \partial^{\nu} h'_{\mu\nu}\ .}
The gauge freedom $\beta^{\mu}{(x)}$ can then be used to set 
\eqn\transverse{
\partial^{\nu} {\bar{h}'_{\mu\nu}} = 0
}
and the remaining equation \munueq\ becomes
\eqn\Lapeq{\sq {\bar{h}'_{\mu\nu}} = 0\ .}
Boundary conditions for this are determined from the 
boundary condition \bcond\ and the gauge shift induced by \bbend, and take the 
form
\eqn\newbc{
\partial_y (h^{\prime}_{\mu\nu} - h^{\prime}\eta_{\mu\nu})_{\vert_{y=0}} = -
{S_{\mu\nu}\over 2M^{d-1}} - 2(\partial_{\mu}\partial_{\nu} -
\eta_{\mu\nu}\partial_{\lambda}^2)\, \alpha^y  \, .
}
In terms of the scalar Neumann Green function $\Delta_{d+1}$ of the
preceding section,
%for the brane background,
%
%\eqn\branegf{
%\sq \Delta_{d+1} (X,X^{\prime}) = {\delta^{d+1}(X - X^{\prime})\over {\sqrt {-
%G}}}
%}
%
%with Neumann boundary conditions
%
%\eqn\branebc{
%\partial_y  \Delta_{d+1} (x,x^{\prime})_{\vert_{y=0}} = 0 \, ,
%}
%
the solution is given by
\eqn\baro{
   \eqalign {h'_{\mu\nu}(X) =
\bar{h}^{\prime}_{\mu\nu} (X) &= -{1\over 2M^{d-1}} \, \
\int \, d^d x^{\prime}
{\sqrt {-g}}\, \Delta_{d+1} (X; x^{\prime},0)\cr
& \biggl[ S_{\mu\nu} (x^{\prime} ) - {1\over d-1} \,
 \eta_{\mu\nu}
S(x^{\prime}) + {1\over d-1} \, {\partial_{\mu}\partial
_{\nu}\over
\partial^2_{\lambda}} \, S (x^{\prime})\biggr]\cr
}
}
where the first equality follows from tracelessness of ${\bar
h}'_{\mu\nu}$; the source on the RHS is clearly 
transverse and traceless as well.
Recall that in this gauge the brane is located at $y=-\alpha^y$.

The quantity $h'_{\mu\nu}$ is appropriate for discussing observations in the
bulk, but a simpler gauge exists for observers 
on the brane.  First note that integration
by parts and translation invariance of $\Delta_{d+1}(x,0;x',0)$ implies
\eqn\exga{
   \eqalign {
{\int d^d x^{\prime}}
{\sqrt {-g}}\, \Delta_{d+1} (x,0;x^{\prime},0 ) \, \
\biggl[ 2 {\partial_{\mu}\partial_{\nu}\over
\partial^2_{\lambda}} - {\eta_{\mu\nu}} {\partial^{2}_{\lambda}
\over \partial^2_{\lambda}} \biggr] S(x') \,  \cr 
= 
(2 \partial_{\mu} \partial_{\nu} - \eta_{\mu\nu} \partial^{2}_{\lambda})
{\int d^d x^{\prime}}
{\sqrt {-g}}\, \Delta_{d+1} (x,0 ; x^{\prime},0 )
{1 \over \partial^{2}_{\lambda}} S(x')\ ; 
}
}
this, together with a gauge transformation using $\beta^\mu(x)$, can be used
to eliminate the third term in \baro.  
We then return to
${\bar h}_{\mu\nu}$ by inverting the gauge transformation \gaugeXm; from the
$d$-dimensional perspective the only gauge non-trivial piece is the
$\alpha^y$ term, which we rewrite using \bbend.  Thus, modulo $d$-dimensional
gauge transformations,
\eqn\branesol{
   \eqalign {
\bar{h}_{\mu\nu} (x) &= -{1\over 2M^{d-1}} \, \
\int \, d^d x^{\prime} \Biggl\{
\, \Delta_{d+1} (x,0;x^{\prime},0) S_{\mu\nu}(x') - \cr
& \eta_{\mu\nu} \biggl[ \Delta_{d+1} 
(x,0;x',0) - {(d-2) \over \Rads} \Delta_{d}(x,x')\biggr]  \,
{S^{\lambda}_{\lambda}(x') \over 2 (d-1)}\Biggr\}\ .
}}
%
%As we'll see in section  to leading order in distance from the
%source, the second and third terms in this expression combine to give a
%simple result that can be interpreted as due to the ``Kaluza-Klein'' modes
%of the graviton.  

Note from \braneGF\ that the zero-mode piece cancels in the term
multiplying $S_\lambda^\lambda$.  Writing the result in terms of the
$d$-dimensional propagator and Kaluza-Klein kernel given in \kkcontrib\
then yields
\eqn\hfinal{\eqalign{ \bar{h}_{\mu\nu} (x)= - {d-2\over 2R M^{d-1}}& \int d^d
x^{\prime} \Delta_d(x,x') S_{\mu\nu}(x')\cr & - {1\over 2M^{d-1}} \int d^d
x^{\prime} \Delta_{KK}(x,x')\left[S_{\mu\nu}(x') - {\eta_{\mu\nu}\over
2(d-1)} S^{\lambda}_{\lambda}(x')\right]\ .}}
The first term is exactly what would be expected from standard
$d$-dimensional gravity, with Planck mass given by\foot{Our conventions are
related to the standard ones for the gravitational coupling $G$ (see {\it
e.g.} \refs{\MTW}) by $M_4^2= 1/16\pi G$ for four dimensions.}
\eqn\Planckm{M_d^{d-2} ={ R M^{d-1}\over d-2 }\ .}
The second term contains the corrections due to the Kaluza-Klein modes.

\subsec{Matter source in bulk}

As a second example of the general solution provided by \munueq, suppose
that the matter source is only in the bulk.  This in particular includes
scenarios with matter distributions on a probe brane embedded at a fixed
$y$ in the bulk\refs{\LyRa}.  

By the Bianchi identities, Einstein's equations are only consistent in the
presence of a conserved stress tensor.  If we wish to consider
matter restricted to a brane at constant $y$, a stabilization 
mechanism\foot{See {\it e.g.} \refs{\GoWi,\DFGK,\CGRT}.}
must be present to support the matter at this constant ``elevation.''
Consider a stress tensor of the form
\eqn\tevst{T_{\mu\nu}= S_{\mu\nu}(x)\delta(y-y_0)}
which is conserved on the brane,
\eqn\brancons{\partial^\mu S_{\mu\nu} = 0\ .}
For simplicity assume $T_{\mu y}=0$.  
Energy conservation in bulk then states
\eqn\econs{\partial_y\left(e^{-dy/R}T_{yy}\right) = -
{e^{(2-d)y/R}\over R} S^\mu_\mu(x)\delta(y-y_0)\ .}
A solution to this with $T_{yy}=0$  for $y>y_0$ is 
\eqn\Tsol{T_{yy} = {e^{[2y_0+d(y-y_0)]/R}\over 
R} \theta(y_0-y) S_\mu^\mu(x)\ .}
We can think of this $T_{yy}$ as arising from whatever physics is
responsible for the stabilization.

Whether we consider matter confined to the brane in this way, or free to
move about in the bulk, the results of
this section give the linearized gravitational solution for a general
conserved bulk stress tensor.  Assuming for simplicity that $T_{\mu \nu}=T_{y
y}=0$ for $y>y_0$, and that $T_{\mu y}\equiv0$, we can gauge fix such that
(see \hpsol,\muysol)
\eqn\gaugeCond{\partial^\mu{\bar h}_{\mu\nu}=0\ ,\ \partial_y h=0\ }
for $y>y_0$.  Thus outside of matter, we see from eq.~\munueq\ that 
$h_{\mu\nu}$ again satisfies the
scalar AdS wave equation.  In particular, for matter concentrated on the
probe brane at $y=y_0$, eq.~\munueq\ gives 
\eqn\munub{ \sq {\bar h}_{\mu\nu}= - {e^{2y_0/R}\over M^{d-1}}
\left[S_{\mu\nu}-{1\over 2(d-1)}\eta_{\mu\nu}
S_\lambda^\lambda\right]\delta(y-y_0) -{1\over M^{d-1}}{\tilde S}_{\mu\nu}\ ,}
where ${\tilde S}_{\mu\nu}$ arises from nonvanishing $\partial^\mu {\bar
h}_{\mu\nu}$ for $y<y_0$ on the RHS of \munueq, resulting from the
stabilization mechanism.  This has solution 
\eqn\Tevsol{\eqalign{
{\bar h}_{\mu\nu}(X) = &-{1\over M^{d-1}} \int d^{d}x' \sqrt{-g}
\Delta_{d+1}(X;x',y_0) e^{2y_0/R} \left[S_{\mu\nu}(x')-{1\over 
2(d-1)}\eta_{\mu\nu}
S_\lambda^\lambda(x')\right]\cr &- {1\over M^{d-1}} \int d^{d+1}X' \sqrt{-G} 
\Delta_{d+1}(X;X') {\tilde S}_{\mu\nu}(X') \ .}}
Again the 
graviton Green function $\Delta_{IJ}^{KL}(X,X')$ is 
given in terms of the scalar Green function $\Delta_{d+1}$ of 
\finalG.

\newsec{Asymptotics and physics of the graviton propagator}

We now turn to 
exploration of  various aspects of the asymptotic behavior
of the propagators given in the preceding two sections, both on and off the
brane.  This will allow us to address questions involving the strength of
gravitational corrections, the shape of black holes, {\it etc.}

\subsec{Source on ``Planck'' brane}

We begin by examining the gravitational field seen on the ``Planck'' brane
by an observer on the same brane.  The relevant linearized field was given
in \hfinal.  This clearly exhibits the expected result from linearized
$d$-dimensional gravity, plus a correction term.  The latter gives a
subleading correction to gravity at long distances.  This can be easily
estimated:  $x\gg R$ corresponds to $qR\ll 1$, where \kkcontrib\
and the small argument
formula for the Hankel functions yields
\eqn\dkkfour{ \Delta_{KK}(x,x') \approx R\int {d^4p\over (2\pi)^4} 
e^{ip(x-x')} \ln(qR) \propto
{R/r^4}}
for $d=4$, with $r=|x-x'|$. 
For $d>4$, we need subleading terms in the expansion of the Hankel
function.  For even $d$,
this takes the general form (neglecting numerical coefficients)
\eqn\Hankexp{H_\nu^{(1)}(x)\sim x^{-\nu}(1+x^2+x^4+\cdots)+x^\nu \ln
x(1+x^2+x^4+\cdots)\ .}
Powers of $q$ in the integrand of \kkcontrib\ 
yield terms smaller than powers of $1/r$ (contact terms, exponentially
supressed terms).  The
leading contribution to the propagator comes from the logarithm, with
coefficient the smallest power of $q$.  This gives
\eqn\dkkmore{ \Delta_{KK}(x,x') \propto \int {d^dp\over
(2\pi)^d}
e^{ip(x-x')}q^{d-4}\ln(qR) \propto
{1/|x-x'|^{2d-4}} }
for $d>4$ and even.  For odd $d$, the logarithm terms are not present in
the expansion \Hankexp, 
and such corrections vanish.  Thus for general even
$d$,
the dominant correction terms are supressed by
a factor of $(R/r)^{d-2}$
relative to the leading term; these are swamped by post-newtonian
corrections.  
Note that in the
special case 
$d=3$, 
$\Delta_{KK}$ was {\it exactly} given by
%
%\eqn\dkkthree{ \Delta_{KK}(x,x') = \ ,}
%
eq.~\kkthree, yielding a correction term of order 
$R/r$ for a static source, as noted in \refs{\EHM}.

One can also examine the short-distance, $r\ll R$, 
behavior of the propagator, which
is governed by the large-$q$ behavior of the Fourier transform.  In this
case we find
\eqn\shortd{ \Delta_{KK}(x,x') \approx \int {d^dp\over (2\pi)^d} 
e^{ip(x-x')} {1\over p} \propto
{1\over |x-x'|^{d-1}}\ .}
Here clearly the Kaluza-Klein term dominates, and gives the
expected $d+1$ dimensional behavior.

Next consider the asymptotics for $z\gg R$ and/or $|x-x'|\gg R$, 
with a source on the Planck brane.   
These 
are dominated by 
the region of the integral with $q\roughly< min(1/z, 1/|x-x'|)$.  This
means that a small argument expansion in $qR$ can be made in the
denominator of the propagator \onbrane, and this gives
\eqn\largez{\eqalign{\Delta_{d+1}(x,z;x',R)\approx &{2\pi i\over R
\Gamma\left({d\over 2}-1\right)} \left({z\over 2}\right)^{d/2}\cr
&\int {d^dp\over (2\pi)^d} e^{ip(x-x')} q^{{d/2}-2} H_{d\over 2}^{(1)}(qz)
\left[1 + {\cal O}((qR)^{d-2} \log qR)\right] \ .}}
In particular $z\gg |x-x'|$ gives $\exp \{ipx\}\approx 1$, and we find  
a falloff
\eqn\zfalloff{\Delta_{d+1}(x,z;x',R) \sim {1\over R
z^{d-2}}}
in the propagator at large $z$.  

Note that this means that 
in the physical case $d=4$, for a
static source, the potential falls like $1/z$ at large $z$.  
This calculation can be taken further to determine the asymptotic shape of
the Green function and potential 
as a function of $r$ and $z$; we do so only for the $d=4$ potential 
though the calculation
may be extended to other $d$.
The static potential for a source at $x'=0$ follows from \largez\ by
integrating over time,
\eqn\statpot{V(x,z)=\int dt \Delta_{4+1}(x,z;0,R)}
which then becomes
\eqn\aspot{V(x,z)\approx  {\pi i z^2\over 2 R} \int {d^3 p\over (2\pi)^3}
e^{ipx}  H_2^{(1)}(ipz)\ .}
This integral is straightforward to perform, and yields 
\eqn\lrzpot{V(x,z) = -{3\over 4\pi} {1\over Rz}\left(1+{2r^2\over 3z^2}\right)
\left(1+{r^2\over z^2}\right)^{-3/2} \left[ 1+ {\cal O} \left({R^2\over
z^2}, {R^2\over r^2} \right)\right]\ ,}
giving the large $r$ and large $z$ dependence.  
We will
return to discuss implications of the $1/z$ behavior shortly.

\subsec{Source in bulk or on probe brane}

Similar results hold for gravitational sources in the  bulk, for
example due to matter on a probe brane as described in \Tevsol.  
As seen there, the detailed field depends on the form of the stabilization
mechanism, but a general understanding follows from consideration
of the scalar propagator entering into \Tevsol.
Begin by considering the 
general scalar propagator \finalG\ for $|x-x'|\gg
R,z,z'$.  This region is again governed by the small-$q$ expansion.
At leading order in $q$, \finalG\ again yields
\eqn\smallq{\Delta_{d+1}(x,z;x,z')\approx {{d-2}\over R} \int {d^dp\over
(2\pi)^d} e^{ip(x-x')} {1\over q^2}
%\left\{ 1+ {\cal O}\left[ (zq)^2,\,
%(z'q)^2,\, (qR)^2\right]\right\}
\ .}
Note that the result is $z$ independent, as in the analogous situation in
Kaluza-Klein theory when we consider sources at large separation compared
to the compact radius; the long-distance field is determined by the zero mode. 

{}From this result we find that the
gravitational potential energy
between two objects at coordinate $z>R$ and 
with $d+1$ dimensional masses $m$ and $m'$ behaves as
\eqn\gravpot{ V(r) \propto 
{1\over R M^{d-1}} 
\left({R\over z}\right)^2
 {mm'\over r^{d-3}} 
%\left[1+{\cal O}\left({R\over r}\right)^2\right]
\ .}
(Note that this potential only includes contributions from the two objects
and not their stabilizing fields.)
This can be rewritten in terms of the $d$-dimensional ``physical'' mass
using
\eqn\dmass{m_d=Rm/z\ .}
%
%This may be alternatively be rewritten in terms of proper distance on the
%probe brane using 
%%
%\eqn\properr{r={z\over R} r_p\ }
%%
%which follows from \Adsmet.
%These powers can be interpreted in $d$-dimensional terms as saying that the
%$d$-dimensional mass of an object of fundamental mass $m$ at $z$ is
%rescaled by a power of $z$:
%
%\eqn\rescalem{ m_d = m \left({R\over z}\right)^{d-2\over 2}\ .}
%
%[ - is this right?].  

For probe-brane scenarios\LyRa, 
it is also important to understand the size of the corrections to this 
formula.  This follows from \finalG\ and the expansions \Hankexp\ and
(again neglecting numerical coefficients)
\eqn\Bessexp{ J_\nu(x)\sim x^\nu(1+x^2+x^4+\cdots)
\ .}
Applying these to the first term in \finalG\ yields (for even $d$) 
\eqn\termone{\eqalign{&{J_{{d\over 2}-1}(qR) \over H_{{d\over 2}-1}(qR) 
}H_{d\over
2}^2(qz)\cr &\sim {R^{d-2}\over z^d} {1\over q^2} \left[1+q^2R^2 + q^2z^2
+ (qR)^{d-2}\ln(qR)+q^dz^d\ln(qz) \right]\left\{1+{\cal O}\left[(qz)^2,
(qR)^2\right]\right\}\ .}}
Here we have dropped subdominant terms.  The corrections involving simple
powers of $q$ again all integrate to yield contact terms at $x=x'$, so the
dominant corrections at finite separation come from the logarithmic terms.
These then give contributions to the propagator of the form
\eqn\propcorro{\Delta(x,z;x',z)\sim {1\over R r^{d-2}} \left[1+ \left({R\over
r}\right)^{d-2} +{z^d\over r^d} +\cdots\right]\ .}
Combining this with a similar analysis of the second term in \finalG\ using
\eqn\termtwo{J_{d\over 2}(qz)H_{d\over 2}(qz) \sim
1+q^2z^2+\cdots+q^dz^d\ln(qz)+\cdots\ }
leads to an expansion of the form
\eqn\propcorr{\Delta(x,z;x',z')\sim {1\over Rr^{d-2}}\left[{1}+ {R^{d-2}\over
r^{d-2}} + {z^d\over r^d} + {z^{2d}\over r^{2d}}{r^{d-2}\over
R^{d-2}}\right] \left[1+{\cal O}\left({z^2\over r^2},
{R^2\over r^2}\right)\right] }
for the propagator for even $d$.  
Which terms provide the dominant correction depends on the magnitude of
$r$.  At long distances,  $r^2> z^d/R^{d-2}$, the first term is the
dominant correction.
In the physical case of $d=4$, the corresponding scale is 
\eqn\crossscal{
r\sim z^2/R \sim
(10^{-4} eV)^{-1}}
in the scenario of \LyRa, and at larger scales the corrections to the
newtonian potential therefore go like $1/r^3$ with a Planck-size
coefficient, and would be swamped by post-newtonian corrections, as with
corrections on the Planck brane.
On the other hand, at shorter scales than \crossscal\ , the last term
 in \propcorr\
is the dominant correction to Newton's law.  
This gives a propagator correction of the form $z^8/r^8$ in
four dimensions.\foot{This is in agreement with \LyRa, and the subleading
corrections can also be obtained from the mode sum.}  
This is the first correction that would be relevant
in high-energy experiments.

It is interesting to investigate the asymptotics of the propagator in more
detail.  Concretely, consider a source at $(x',z')$ with $z'\gg R$. In the
limit $r^{d+2}$ and/or $z^{d+2}\gg (z')^{2d}/R^{d-2}$, or $z^{2d} \ll 
(z')^{d+2} R^{d-2}$, the
first term in the propagator \finalG\ dominates over the bulk AdS part
given by the second term.  Consequently, the behavior is given by 
expressions like \largez\ and \lrzpot\
(where $z$ is replaced by $z'$ for the latter limit). 
On the other hand,
for $r^{d+2},z^{d+2} \ll  (z')^{2d}/R^{d-2}$, the bulk 
AdS portion dominates and hence
determines the shape of the potential. 
This AdS
propagator is explicitly given in terms of a hypergeometric
function\refs{\BuLu,\InOo}. 
This transition to bulk AdS behavior is that
seen in \propcorr.
Indeed, the asymptotic behavior of
the bulk AdS propagator for $r\gg z,z',R$,
\eqn\adsgfa{ G_{\rm AdS}\sim \left[{zz'\over r^2 + (z-z')^2}\right]^d\ ,}
is what determines the $z^{2d}/r^{2d}$ corrections discussed there.
%the behavior of the near-fields as
%well; concretely, consider a source at $(x',z')$ with $z'\gg R$.  The large
%$r$ behavior is therefore given in \smallq.  Another region of interest is
%$z' \gg z$.  In this limit, with field point near the Planck brane, the
%potential is again given by expressions like \largerz, \lrzpot\ with primed
%and unprimed variables exchanged.  Next,
%consider the limit $r\sim z\sim z' \gg R$.  Here the second term in the
%Green function \finalG, corresponding to the bulk AdS propagator,
%dominates, as may easily be checked by a scaling argument.  The AdS
%propagator is explicitly given in terms of a hypergeometric
%function\refs{\BuLu,\InOo}, which determines the shape of the potential in
%this region.  Note also that it is precisely the asymptotic behavior of
%this bulk AdS
%propogator for $r\gg z,z',R$,
%%
%\eqn\adsgfa{ G_{\rm AdS}\sim \left[{zz'\over r^2 + (z-z')^2}\right]^d\ .}
%%
%that determines the $z^8/r^8$ corrections discussed above.  

Finally, the short-distance 
bulk asymptotics are also easily examined.  Specifically, let 
$|x-x'| \ll R$ and
$|z-z'|\ll R$; again, 
the result is governed by the large $qR$ behavior of the Green
function.  From \finalG\ we find 
\eqn\smallzxn{ \Delta_{d+1}(x,z;x',z')\approx -{i\pi\over 4 R^{d-1}}
(zz')^{d\over 2}  \int {d^d p\over (2\pi)^d} e^{ip(x-x')} H_{d\over
2}^{(1)}(qz_>)  H_{d\over 2}^{(2)}(qz_<)\ ,}
where $H^{(2)}=J-iY$.  Aside from non-standard boundary conditions, this is
the usual propagator for anti-de Sitter space, and at short distances
compared to $R$ will reduce to the flat space propagator 
(see {\it e.g} \refs{\FSS}).  In particular, suppose that $|x-x'|\ll
z,z'$.  Then we may also use the large $qz$ expansion, which gives
\eqn\smallzx{\Delta_{d+1}(x,z;x',z') \approx -{i\over 2}
\left({zz'\over R^2}\right)^{d-1\over 2} 
\int {d^dp\over (2\pi)^d} e^{ip(x-x')+ iq(z_>-z_<)}{1\over q}\ ,}
resulting in
\eqn\smalldx{\Delta_{d+1}(x,z;x',z') \sim \left({zz'/R^2 \over |x-x'|^2 +
|z-z'|^2}\right)^{d-1 \over 2} }
which is the expected behavior in $d+1$ dimensions.

\subsec{Off-brane profile of gravitational fields and black holes}

These results can readily be applied to discuss some interesting properties
of black holes and more general gravitational fields in the context of 
brane-localized gravity.  In particular, one might
ask what a black hole -- or more general gravitational field --
formed from collapsing matter on the brane looks like.  In a na\"\i ve
analysis, considering only fluctuations in the zero mode \hzm, 
one finds that metrics
of the form 
\eqn\ricmet{ds^2= {R^2\over z^2} [ dz^2 + g_{\mu\nu}(x) dx^\mu dx^\nu] }
are solutions of Einstein's equations for general Ricci flat
four-dimensional metric $g_{\mu\nu}(x)$.  These solutions are, however,
singular on the horizon at $z=\infty$, as was discussed in \CHR.
Nonsingular solutions require excitation of the other modes of the
graviton.  
Although exact
solutions are elusive except for $d=3$ \EHM, 
the linear analysis of this paper gives us the
general picture.  

Consider a massive object, $m\gg 1/R$, 
on the Planck brane.  Without loss of generality its
metric may be put in the form \metfluct.  
The linear approximation is valid
far from the object (for more discussion, see the next subsection).  As
we've seen in section 4.1, at long distances along the brane we recover
standard linearized gravity, with potential of the form \gravpot.
Transverse to the brane, we use the expression \zfalloff, which, with
\baro, implies that 
\eqn\hooz{ h_{00}\sim {m\over M_d^{d-2} z^{d-3}}}
(the extra power of $z$ arises because we are considering the static
potential).  Note that the
proper distance off the brane is given by $y$, defined in eq.~\ydef.  Thus
in fact the metric due to an object on the brane 
falls exponentially with proper distance off the
brane.  

If the mass is compact enough to
form a black hole, the corresponding horizon is the surface where (for
static black holes) $h_{00}(x,y) = 1$.    In the
absence of an exact solution, the horizon is approximately characterized by
the condition $h_{00}\sim 1$.  Since we recover linearized $d$-dimensional
gravity at long distances along the brane, 
the horizon location on the brane is given by the 
usual condition in terms of the
$d$-dimensional Planck mass \Planckm:
\eqn\horr{r_h^{d-3}\sim {m\over M_d^{d-2}}\ .}
Transverse to the brane
%, we use the expression \zfalloff, which, with
%\baro, implies that 
%%
%\eqn\hooz{ h_{00}\sim {m\over M_d^{d-2} z^{d-3}}}
%%
%(the extra power of $z$ arises because we are considering the static
%potential).  
\hooz\ implies that the 
horizon is at $z_h\sim r_h$.  
Taking into account the exponential relation between $y$ and $z$, 
a rough measure of the proper ``size'' of the black hole transverse to the 
brane
is 
\eqn\hory{ y_h\sim R \log\left[ \left({m\over M_d^{d-2}}\right)^{1\over d-3}
{1\over R} \right]\ .}
So while the size along the brane grows like $m^{1/d-3}$, the thickness
transverse to the brane grows only like $\log m$.  The black hole is shaped
like a pancake.\foot{The value of $z_h$ was guessed by \refs{\CHR} who
nonetheless referred to the resulting object as a black cigar instead of a
black pancake.  Ref.~\EHM\ also independently found the black pancake picture 
in the
special case of $d=3$, where they were able to find an exact solution.}
This black pancake has a constant coordinate radius $r_h$ in $x$ coordinates 
for
$z\roughly< z_h$, as seen in \smallq.  From \Adsmet\ we then see that the
proper physical size scales with $z$ as
%One might also ask how the black hole appears
%to an observer on a probe brane.  This brane will 
%intersect the black hole as long as it is at
%$z\roughly< z_h$.  From \smallq\ and \Adsmet\  we see that the 
%coordinate size of the horizon on the probe brane is still $r_h$, but that
%the proper size will be rescaled by $R/z$:
%
\eqn\zhorr{r_h(z) = Rr_h/z\ .}

An amusing consequence of this picture of black holes 
is the possibility that matter
on a collision course with a black hole from the four-dimensional
perspective can in fact pass around it
through the fifth dimension.\foot{We thank L. Susskind for 
this observation.}  This suggests that from the perspective of the 
four-dimensional
observer, matter can pass through a black hole!  To investigate this more
closely, consider the specific case of a 
black hole on the Planck brane.  A bulk mode
(graviton or other bulk matter) can bypass the black hole by traveling 
at $z>z_h$.
Suppose first that the proper wavelength of the bulk mode is $\lambda>R$.
>From the four dimensional perspective of the Planck brane the
wavelength is redshifted,
\eqn\rshift{\lambda_{d} \sim \left({z\over R}\right) \lambda > z_h
\left({\lambda\over R}\right)  \ .}
Thus for $\lambda>R$, such matter has a wavelength larger than the size of
the black hole, which suggests that this process is difficult to
distinguish from passage around the black hole in four dimensions.
Next consider a mode with wavelength
$\lambda<R$.   Such a perturbation obeys the geometric optics
approximation and will fall into the horizon at $z=\infty$; if emitted from
the Planck brane it will never return.  
However, although this particle is moving in the $z$ direction
it might be possible that a 
four-dimensional observer could observe it's projected gravitational field 
emerging
from the far side of the black hole.  We leave further investigation of
this process for future work.

A semi-quantitative picture
of the shape of gravitational fields due to sources in the bulk, e.g. on
a probe brane, also follows from the asymptotics of the propagator as described
in section 4.2.  Again, let the source be at $x'=0$ and $z'\gg R$.  At short
distances
$r^2,z^2 \ll (z')^d/R^{d-2}$, the shape is given by the bulk AdS
propagator (which, as expected, for $r\ll R$ and $z-z' \ll R$ reduces to
the flat $d$+1 dimensional result as seen in \smalldx.)
%the field is effectively $d$+1
%dimensional as shown in \smalldx, and in 
%particular small black holes will look like those in
%$d$+1-dimensional flat space.   
At longer distances, 
$r^2$ and/or $z^2\gg (z')^d/R^{d-2}$, or $z^d \ll (z')^2 R^{d-2}$, the zero
mode begins to dominate with a shape given by 
\lrzpot\ or its $d$-dimensional generalization.  This determines the
pancake shape discussed above.

\subsec{Strength of perturbations}

Lastly we turn to the question of consistency of both the linear
approximation, as well as of the scenario of $\RaSu$ within the context of
a complete theory of quantum gravity, such as string theory (see the next
section for comments on the latter).  For simplicity we confine the
discussion to the $d=4$ case, although the results clearly extend.

If one were to base one's analysis solely on the properties of the
zero-mode of \hzm, serious questions of consistency immediately arise.  For
one thing, the black tube metrics mentioned in the introduction become
singular at the AdS horizon\refs{\CHR}. 
Alternately, trouble would be encountered when one considered consistency of
linearized gravity, or of the overall scenario after higher-order corrections
to gravity of the form
\eqn\metshift{S\rightarrow S+\alpha\int d^5X \sqrtG \calR^2}
are taken into account.  The apparent trouble occurs due to the growth of
the curvature of the zero-mode correction to the metric, \gravfluct.

To see the problems more directly, recall that the general structure
of the curvature scalar is 
\eqn\curvexp{\calR\sim G^{-2} \partial^2 G\sim z^2\left(\partial^2 h +
h\partial^2 h + h^2 \partial^2 h + \cdots\right)\ .}
Similarly, scalars comprised of $p$ powers of the curvature will grow like
$z^{2p}$.  According to this na\"\i ve analysis, such terms would dominate
the action of the zero mode, 
which would suggest the scenario is inconsistent.
  Likewise, consider the Einstein-Hilbert action,
\eqn\zmact{S\sim \int d^5X {1\over z^3} \left(
h\partial^2 h + h^2 \partial^2 h + \cdots\right)\ .}
Rescaling $h\rightarrow z^{3/2}h$ gives an expansion of the form
\eqn\rescact{S\sim \int d^5X \left(h\partial^2 
h + z^{3/2}h^2 \partial^2 h + \cdots\right)\ }
with diverging expansion parameter, a similar problem.  
What these arguments do not account
for is that far from the brane the zero-mode no longer dominates, and in
fact for a static source the perturbation falls as $h\sim 1/z$ as seen 
in \hooz.  This
ensures that $\calR^p$ corrections and non-linear corrections to linearized
Einstein-Hilbert action are indeed suppressed, in accord with the intuition
that a localized gravitational source produces a localized field, not a
field that is strong at the horizon.

This addresses the question of potential strong gravitational effects near
the horizon in the static case. We see that the source of the apparently
singular results is the treatment of the zero mode in isolation; with the
full propagator the field dies off as it should with $z$.  This leaves open
questions about potentially strong effects in dynamical processes. However,
once again given the falloff of the propagator with $z$, we expect this
situation to be very similar to that of scattering in the presence of a
black hole, where for most local physics processes ( {\it e.g.}  near the
brane), the existence of the black hole is irrelevant.  However, while we
have not attempted to fully describe particles that fall into the AdS
horizon, and radiation that is potentially reemitted from the horizon, this
may require confronting the usual puzzles of strongly coupled
gravity.\foot{For related comments see
\WittITP.}

\newsec{Relation to the AdS/CFT correspondence}

Treatment of the scenario of \RaSu\ within string theory in the special
case where the bulk space is AdS may naturally
incorporate the AdS/CFT correspondence 
as \refs{\Hver,\MaldUnp,\WittITP,\Gubs} have
recently advocated, although the analogous statement is not 
known for the case
of more general bulk spaces.  
%For followup discussion and an interesting
%application, see \refs{\Gubs}.}  
Here we will make some comments on this and on the
connection with our results.

Quantization of the system in the 
scenario of \RaSu\ would be performed
by doing functional integrals -- or whatever ultimately replaces
them in string theory -- of the form
\eqn\gfcnint{\int \cald \Psi \int \cald G e^{iS}}
over configurations near the background \adsmet; in this section we will
take the brane to reside at a general radius $z=\rho$ and treat it as
a boundary.  Here the action
is of the form \dact, plus the required surface term $2M^3\int_\rho d^4x \sqrtg
K$ in the presence of the boundary, and $\Psi$ represents
all the non-gravitational  ``matter'' fields of string theory, including those
describing the matter moving on the brane.  In this section we work with
$d=4$, though generalization is straightforward.

Consider the metric part of this integral, 
\eqn\genfun{Z[\Psi]= \int \cald G e^{iS}}
with fixed matter background
fields.  One may think of doing this functional integral in two steps:
first one integrates over all metrics that match a given boundary metric
$g_{\mu\nu}$ on the brane, and then one integrates over all such boundary
metrics.  The latter integral enforces the boundary condition given in
\bcond, which in general can be written in terms of the extrinsic curvature
$K_{\mu\nu}$ of the brane as 
\eqn\extcond{(K_\nu^\mu-\delta_\nu^\mu K)_{|z=\rho} = S_\nu^\mu/4M^3\ .}
If one in particular works in the leading semiclassical approximation, this
functional integral takes the form
\eqn\Zappr{Z[\Psi]\approx e^{{i\over 4M^3}\int dV dV'  T_{IJ}(X) 
\Delta^{IJ,KL}(X,X') T_{KL}(X')}\ }
Here $T_{IJ}$ may represent a source either in the bulk or on the
brane.  $\Delta^{IJ,KL}$ is the gravitational propagator derived in
section three; in particular, as we've seen it obeys the gravitational
analog of Neumann boundary conditions.

For sake of illustration, consider the case where the background fields
$\Psi$ are only turned on at the brane; also, for simplicity we will
ignore fluctuations of matter fields in the bulk.
In line with the above comments, we may rewrite 
\eqn\adsconn{Z[\Psi]= \int Dg  e^{{i\over 2}\int d^4x \sqrtg(\calL_{\rm
brane}(\Psi) -\tau) } Z[g,\rho]}
where $Dg$ represents the integral over boundary metrics,  
the bulk functional integral
\eqn\zdef{Z[g,\rho]=
\int_{G_{|z=\rho}=g}  \cald G e^{i\int d^5X \sqrtG (M^3 \calR-\Lambda)
 +2iM^3 \int d^4x
\sqrtg K}}
is performed with fixed (Dirichlet)
boundary conditions as indicated, and the factor of 1/2 in \adsconn\ 
results from
orbifolding to restrict to one copy of AdS instead of two copies on
opposite sides of the brane. 
The latter functional plays a central
role in the AdS/CFT correspondence\refs{\Mald}, which
states\refs{\GKP,\WittAds}
\eqn\adsone{\lim_{\rho\rightarrow0} Z[g,\rho] = \biggl\langle e^{i\int
g_{\mu\nu}T^{\mu\nu} } \biggr\rangle_{\rm CFT}\ .}
Actually, as it stands, this equation is not quite correct -- counterterms
must be added to regulate the result as $\rho\rightarrow0$.  The
infinite part of these counterterms were worked out in
\refs{\HeSk,\BaKr}; including them, an improved version of the
correspondence is 
\eqn\adstwo{  \lim_{\rho\rightarrow0} e^{i\int d^4 x\sqrtg (b_0 + b_2 \calR +
b_4 \log(\rho)\calR_2 )} Z[g,\rho]  = 
\biggl\langle e^{i\int
g_{\mu\nu}T^{\mu\nu} } \biggr\rangle_{\rm CFT}}
where $b_0={-6M^3/R}$, $b_2=-RM^3/2$, $b_4=2M^3R^3$, and the
RHS is given in terms of renormalized quantities,
 and 
\eqn\rsqdef{\calR_2 = -{1\over 8} \calR_{\mu\nu} \calR^{\mu\nu} + {1\over
24} \calR^2\ .}

In the present context, a version of the AdS/CFT correspondence for finite
$\rho$ is needed.  It is natural to conjecture that 
\eqn\finads{ e^{i\int d^4 x\sqrtg (b_0 + b_2 \calR +
b_4 \log(\rho)\calR_2 )} Z[g,\rho]  = 
\biggl\langle e^{i\int
g_{\mu\nu}T^{\mu\nu} } \biggr\rangle_{\rm CFT, \rho}}
where $\rho$ defines a corresponding cutoff for the conformal field 
theory.\foot{One may also expect that finite counterterms could be added by
shifting the $b_i$s by terms that vanish as $\rho\rightarrow0$,
though the diffeomorphism-invariant origin of these terms is
subtle except for $b_4$. There may also be counterterms involving higher
powers of $\calR$.  We thank V. Balasubramanian and P. Kraus 
for a discussion on
this point.}  Related ideas have appeared in a discussion of the
holographic renormalization group\refs{\BaKrHol}.
We can think of this as providing a definition of the generating functional
$Z[g,\rho]$ in terms of that of the cutoff conformal field theory, and thus
can rewrite \adsconn\ as
\eqn\adsdef{Z[\Psi]= \int Dg  e^{i\int d^4x \sqrtg[{1\over 2}\calL_{\rm
brane}(\Psi) -(\tau/2 +b_0) - b_2 \calR -
b_4 \log(\rho)\calR_2 ]} \biggl\langle e^{i\int
g_{\mu\nu}T^{\mu\nu} } \biggr\rangle_{\rm CFT, \rho}\ .}

The semiclassical/large-N approximation to this may then be compared to
\Zappr.  The parameters $\tau$ and $b_0$ cancel. 
The curvature term is responsible for the 4-dimensional
part of the graviton propagator that was seen in \propsep, \hfinal.  The
curvature-squared term and the CFT correlators clearly correct this result
-- from the bulk perspective these corrections arise from the  bulk modes which
have been integrated out.   In
particular, the two-point function of the CFT stress tensor,
\eqn\twopoi{\langle T^{\kappa\lambda}  T^{\mu\nu} \rangle_{\rm CFT, \rho}  =  
-{\delta^2\over \delta g_{\kappa\lambda} 
\delta g_{\mu\nu}} \biggl\langle e^{i\int
g_{\mu\nu}T^{\mu\nu} } \biggr\rangle_{\rm CFT, \rho}}
gives a leading contribution to the propagator, as argued by
Witten\WittITP, and as shown in fig.~3.  One may readily check that the
form of the corrections agrees.  From \GKP, $\langle TT(p)\rangle \sim p^4
\log p$.  Attaching two external gravitons gives an extra $1/p^4$,
resulting in a correction
\eqn\deltatt{\Delta_{TT}\sim \int d^4p e^{ipx} \log p \sim {1\over r^4}\ ,}
in agreement with \dkkfour.
A more careful computation could be made to check the
coefficient.  However, from the above connection, we see that this is not
an independent check of the AdS/CFT correspondence:  $\langle TT\rangle $
was computed in \GKP\ from supergravity 
precisely by using the formula \adsone\ and
regulating, which is simply a rearrangement of the steps in the above
discussion.  One may in fact relate higher-order terms in the momentum in the
full propagator of \braneGF\ to diagrams with multiple insertions of 
$\langle TT(p)\rangle_\rho$ and counterterms.  An outline of a general
argument for this follows from two alternative ways to derive \Zappr\ from 
\genfun.
On one hand, we could first integrate over the boundary metrics
$g$, finding the constraint \extcond.  When we next integrate over 
the bulk metrics $G$,
we obtain \Zappr\
with the propagator obeying the analog of Neumann boundary
conditions.  On the other hand, we could first integrate over the bulk
metrics.  This will lead to an effective action for $g$; according to 
the AdS/CFT prescription \adsdef\ the kinetic operator in the quadratic term
in this action is shifted by $\langle TT(p)\rangle_\rho$.
Then integrating over the boundary metric $g$ yields \Zappr, where the
propagator is the inverse of this shifted kinetic operator.

\ifig{\Fig\Three}{This diagram represents corrections to the graviton
propagator due to the cutoff conformal field theory.}{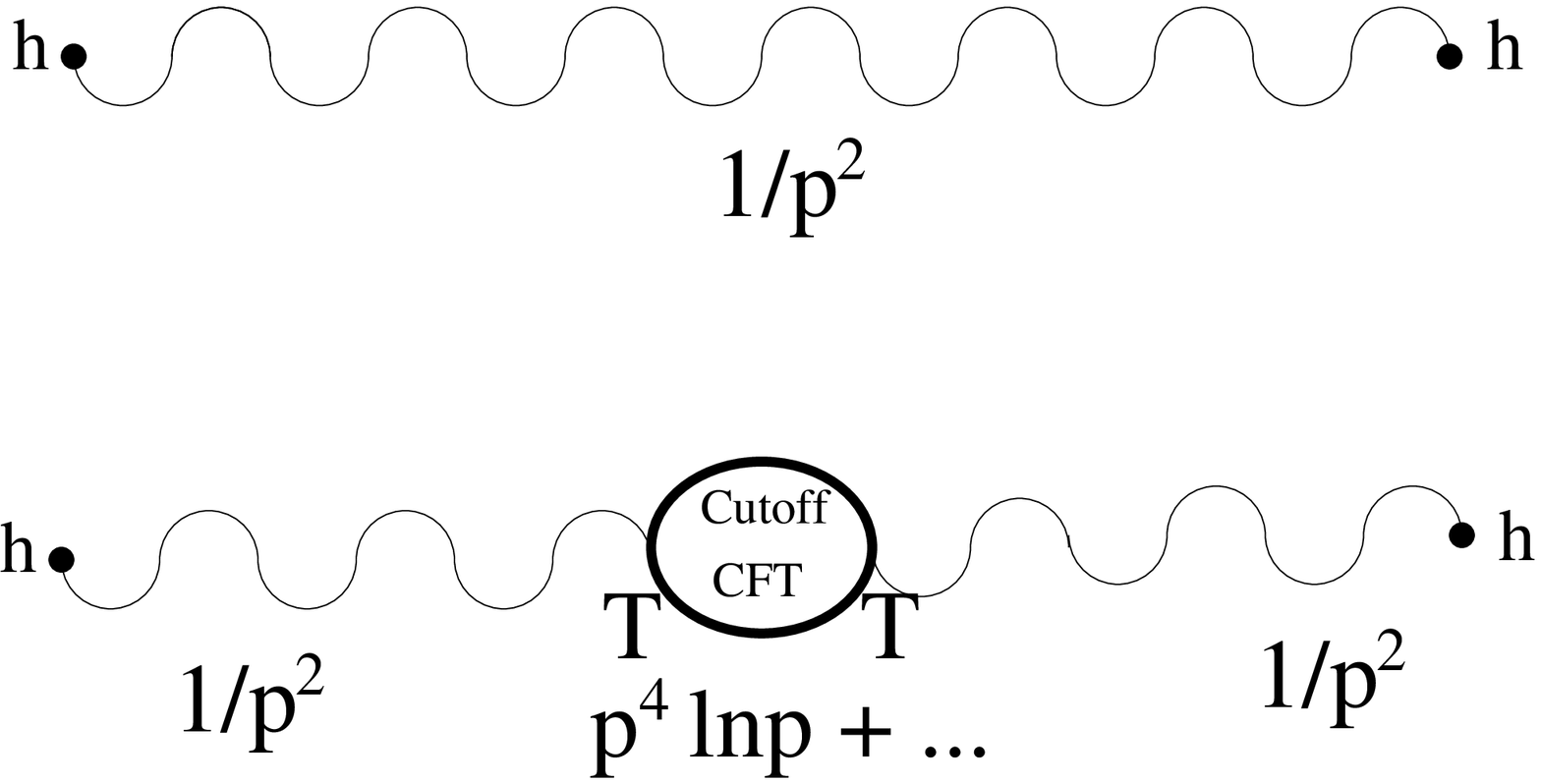}{2.0}

The suggested connection with the AdS/CFT correspondence
provides a potentially interesting new interpretation 
for the scenario of \RaSu, since the bulk
may be completely replaced by the 4d cutoff conformal field theory.  
There are,
however, non-trivial issues in defining the regulated version of this
theory, and furthermore for large radius the gravitational description is
apparently a more useful approach to computation,
since the 't Hooft coupling is large.
Notice also that in a sense this is not
strictly an induced gravity scenario, since the CFT only
induces the corrections due to the bulk modes;  eq.~\adsdef\ includes a
separate four-dimensional curvature term.

\newsec{Conclusion}

The scenario posed in \RaSu\ has by now survived several consistency
checks, including those of this paper.  In a linearized analysis, this
paper has outlined many interesting features of gravity and gravitational
corrections in this scenario.  These await a treatment in an
exact non-linear analysis.  Many other questions remain, including other
aspects of phenomenology and cosmology, and that of a  first-principles
derivation of such scenarios in string theory.

\bigskip\bigskip\centerline{{\bf Acknowledgements}}\nobreak

We thank V. Balasubramanian, O. DeWolfe, 
S. Dimopoulos, D. Freedman, M. Gremm, 
S. Gubser, G. Horowitz, I. Klebanov, R. Myers, 
J. Polchinski, L. Rastelli, R. Sundrum, L. Susskind, H. Verlinde, and E. Witten
for many useful discussions.  The work of A. Katz and L. Randall was
supported in part by DOE under cooperative agreement DE-FC02-94ER40818 and
under grant number DE-FG02-91ER4071.  That of S. Giddings was supported in
part by DOE contract number DE-FG-03-91ER40618.  We also thank the ITP at UCSB
for its hospitality while this  research was being performed.

\listrefs
%\listfigs
\end